# A Comparative Study of Interface Techniques for Transmission and Distribution Dynamic Co-Simulation

Qiuhua Huang, Renke Huang, Rui Fan, Jason Fuller,
Trevor Hardy, Zhenyu (Henry) Huang, *Fellow, IEEE*
Pacific Northwest National Laboratory
Richland, WA, USA
{qiuhua.huang|renke.huang|rui.fan|jason.fuller|trevor.hardy|zhenyu.huang}@pnnl.gov

Vijay Vittal, *Fellow, IEEE*
Arizona State University
Tempe, AZ, USA
vijay.vittal@asu.edu

*Abstract*— Transmission and distribution (T&D) dynamic co-simulation is a practical and effective approach to leverage existing simulation tools for transmission and distribution systems to simulate dynamic stability and performance of T&D systems in a systematic manner. Given that these tools are developed as stand-alone programs and there are inherent differences between them, interface techniques become critical to "bridge" them. Two important unsolved questions are: 1) which interface technique is better and should be used, and 2) how the modeling and simulation capabilities in these tools that are available and can be exploited for co-simulation should be considered when selecting an interface technique. To address these questions, this paper presents a comparative study for different interface techniques that can be employed for T&D dynamic co-simulation. The study provides insights into the pros and cons of each interface technique, and helps researchers make informed decisions on choosing the interface techniques.

## I. INTRODUCTION

There are increased interactions between the transmission and distribution (T&D) systems in many power systems in recent years, largely due to widespread adoption of distributed energy resources (DER). As this trend is expected to continue, utilities will need to plan and operate T&D system in a more coordinated or even integrated manner [1]. In this context, adequate simulation tools for supporting T&D system planning and operation is needed [2]-[4]. However, only a few phasor domain simulation tools and electromagnetic transient (EMT) simulation tools support T&D system modeling and simulation. Most existing simulation tools adopted for system planning and operation are designed for either transmission or distribution systems only. T&D co-simulation can be an effective solution to meet the need, mainly because it leverages existing tools, models and simulation data, and avoids the risks and cost of developing new tools and simulation data sets [2].

Given the differences in the modeling and simulation methods between transmission and distribution system simulators, techniques for accurately and efficiently interfacing simulators of both domains are generally recognized as the most challenging part of developing T&D co-simulation. Increasing efforts have been devoted to this area in recent years. Several frameworks, including Framework of Network Co-Simulation (FNCS) [5], HELICS [6] have been developed to facilitate co-simulation for power systems, including T&D co-simulation. These are general co-simulation frameworks and they facilitate communication, data exchange and time synchronization for co-simulation. With these frameworks, users still need to select and develop proper interface techniques for connecting different simulators.

Previous research efforts were mainly focusing on developing interfaces for steady-state and quasi-steady-state applications such as power flow (PF) [5] and market solution [4], whereas relatively few efforts has been directed towards T&D dynamic co-simulation. In the authors' previous research [2]-[3], three-sequence and three-phase phasor representations were used for modeling transmission and distribution systems, respectively. When transmission and distribution simulators are interfaced, multi-area Thévenin equivalent (MATE) was adopted as the interface model (IM), and an iterative, parallel interaction scheme (IS) was used in [2]. In [3], the interface models for T&D dynamic co-simulation are 3-phase total loads (for representing distribution systems in the transmission system) and voltage source (for representing the transmission system in distribution systems). In addition, a non-iterative, parallel interaction scheme was used in [3], consequently a small time step was used to contain the errors. It should be noted that both implementations discussed above leveraged the openness and flexibility of open source software. The same modeling and interfacing techniques may not be feasible or practical for other third-party or commercial simulation tools. Both implementations, along with other existing and potential interface techniques, have not been directly and transparently compared. Thus, it is important to comprehensively investigate different interface techniques and identify their performance, applicability and limitations.

This paper fills the gap by comprehensively comparing several interface techniques, including interface models and interaction schemes, through theoretical analysis and numerical simulation. The main contributions of this paper include: 1) providing comprehensive comparison results for different interface techniques; 2) helping researchers and engineers make informed decisions on choosing a proper interface technique. The remainder of the paper is organized as follows:

This work was supported by the U.S. Department of Energy (DOE) Grid Modernization Laboratory Consortium, as part of the project titled "Development of Integrated Transmission, Distribution, and Communication (TDC) Models" (GMLC 1.4.15). Pacific Northwest National Laboratory is operated by Battelle for the DOE under Contract DE-AC05-76RL01830.

TABLE I. Interface Models for T&D Dynamic Co-Simulation Considering Different T&D System Modeling Approaches and Algorithms

| Num. | T&D System Modeling | T&D Simulation Algorithm | Interface Models | Key assumptions | Limitations |
|---|---|---|---|---|---|
| 1 | T: positive-sequence; D: three-phase | T: positive-sequence TS; D: three-phase PF | T→D: three-phase balanced voltage source; D→T: positive-sequence equivalent load | Unbalanced conditions, if any, only occur in D, no unbalance in T; T&D boundary conditions are balanced | Unbalanced fault can't be applied in T; no fault is allowed in D |
| 2 | | T: positive-sequence TS; D: three-phase DS | T→D: three-phase balanced voltage source; D→T: positive-sequence equivalent load | Unbalanced conditions, if any, only occur in D, no unbalance in T; T&D boundary conditions are balanced | Unbalanced fault can't be applied in T |
| 3 | | T: positive-sequence TS; D: three-phase DS | T→D: three-phase Thevenin equivalent; D→T: positive-sequence equivalent load | Unbalanced conditions, if any, only occur in D, no unbalance in T; T&D boundary conditions are balanced. | Unbalanced fault can't be applied in T |
| 4 | T: three-sequence; D: three-phase | T: three-sequence TS; D: three-phase PF | T→D: three-phase voltage source; D→T: three-sequence equivalent load | While T is physically balanced, impacts of unbalanced conditions in D can be captured in T; T&D boundary conditions can be unbalanced. | No fault is allowed in D |
| 5 | | T: three-sequence TS; D: three-phase DS | T→D: three-phase voltage source; D→T: three-sequence equivalent load [3] | While T is physically balanced, impact of unbalanced conditions in D can be captured in T; T&D boundary conditions can be unbalanced. | —— |
| 6 | | T: three-sequence TS; D: three-phase DS | T→D: three-phase Thevenin equivalent; D→T: three-sequence equivalent load | While T is physically balanced, impact of unbalanced conditions in D can be captured in T; T&D boundary conditions can be unbalanced. | Unbalanced fault can't be applied in T |
| 7 | | T: three-sequence TS; D: three-phase DS | T&D: three-sequence multi-area Thévenin equivalent (MATE) [2] | While T is physically balanced, impact of unbalanced conditions in D can be captured in T; T&D boundary conditions can be unbalanced. | —— |

Note: DS denotes dynamic simulation; PF denotes power flow; TS denotes transient stability; "T→D" denotes representing the transmission system in the distribution systems during co-simulation; "D→T" denotes representing the distribution systems in the transmission system during co-simulation

firstly, several interface techniques considering the practical modeling and simulation approaches for T&D systems are proposed and presented in section II. Test cases and simulation results are presented in Section III. Conclusions are drawn and future research directions are suggested in Section IV.

## II. INTERFACE TECHNIQUES FOR T&D SYSTEM DYNAMIC CO-SIMULATION

### A. T&D System Modeling and Dynamic Simulation

Traditionally, transmission and distribution systems are generally modeled differently. The differences have to be carefully taken into account when selecting and developing interface techniques for T&D system dynamic co-simulation. Considering the modeling assumptions and capabilities commonly found in existing phasor domain transmission and distribution simulators, two modeling approach combinations for T&D systems are identified and summarized in Table I.

It should be noted that three-sequence modeling capabilities are actually available in many existing transmission simulation tools, but not commonly used in electromechanical transient stability (TS) simulation. Traditionally, TS simulation is based on positive-sequence (single-phase) modeling, while three-sequence modeling are widely used in the short-circuit or fault analysis modules. As shown in [2], three-sequence TS simulation algorithm can be developed by combing the conventional positive-sequence TS simulation with the negative- and zero-sequence network solution functions available in the short-circuit module.

With three-sequence modeling and TS simulation capabilities, unbalanced conditions in the transmission system can be better captured. This paper will study to what extent these three-sequence modeling and TS simulation capabilities contribute to the T&D co-simulation accuracy, by comparing with the positive-sequence counterparts. The results are expected to help other researchers and developers make better informed decision regarding which modeling and simulation approaches to adopt for the transmission systems, if options are available.

Given that not all the distribution system simulation tools have full-scale dynamic simulation (DS) capability, power flow algorithms have been leveraged as the network solution to realized distribution system dynamic simulation in some previous research. Users usually have to develop dynamic models for components such as induction motors, and interface them with the distribution system network (usually in the form of loads or current injections). During dynamic simulation, these dynamic models obtain the distribution network states after the power flow is solved. One main concern is the convergence issue of the power flow solution under low voltage or system fault conditions.

### B. Interface Models

An IM is the representation and the associated data published by or extracted from a simulator and shared with other simulators to achieve co-simulation. A good IM should not only adequately represent the effects of the system simulated by one simulator in other systems simulated by other simulator(s), but also be compatible with the system representation and simulation algorithm in other simulator(s). The general criteria for selecting IMs for T&D dynamic co-simulation are as follows:

a) The selected IMs should be compatible with the T&D system modeling representations and the simulation algorithms;
b) Availability of the information required for calculating the IMs;
c) Robustness of the IMs for co-simulation, particularly under disturbance conditions.

With these considerations, seven IMs (shown in Table I) are selected for investigation in this paper. They are selected primarily based on criterion a). The associated T&D system modeling and simulation algorithm for each IM are also shown in Table I. It should be noted that, except for the interface model #7, equivalent loads are used to represent distribution systems in the transmission system, whereas Thévenin equivalent (either positve-sequence or three-sequence) is only adopted for representing the transmission system in the distribution systems, but not the opposite. There are two main reasons for this: 1) when the transmissoin system is modeled in positive-sequence, there are considerable accuracy concerns of using positive-sequence Thévenin equivalent representation for a three-phase unbalanced distribution system; 2) for a three-phase unbalanced distribution system, the derived three-sequence Thévenin equivalent is coupled among three sequences, which makes it incompatible with the three-sequence decoupled modeling for the transmission system.

*1) Voltage source and Thévenin equivalent for representing transmission systems in distribtuion systems:* Voltage sources are widely used model for representing the transmission system in distribution systems, particularly for steady-state applications, mainly because they are simple and can be directly obtained from the TS simulation results. The main issue with voltage sources as IMs is that the transmission system electrical strength at the T&D boundary cannot be acculately represented. In contrast, a Thévenin equivalent can better reflect the transmission system strength, which is critical for simulating cases with faults applied to the distribtuion systems. On the other hand, Thévenin equivalents cannot be directly extracted from the TS simulation results. Thus, some additional calculation or extension is required for the transmission system simulator to calculate the Thévenin equivalents. When the transmissoin system is modeled by only the positive-sequence, the Thévenin equivalent is derived from positve-sequence information and thus is three-phase balanced. For a three-sequence modeling transmission system, the Thévenin equivalent is derived using the three-sequence networks, and thus could be unbalanced. Futhermore, it should be noted that voltage source does not directly fit the network solution formulation (with the *YV=I* form) of dynamic simulation, while Thévenin equivalent can be converted to Norton equivalent and added to the existing formulation. In this paper, to circumvent this issue, a voltage source has to be treated as a Thévein equivalent with a very small impedance.

*2) Positive-sequence and three-sequence equivalent loads for representing distribution systems in transmission systems:* Equivalent loads are usually adopted for representing distribution systems in transmission systems, mainly because the equivalent load values can be easily obtained. In this paper, equivalent loads are in the form of constant current, because current injections can be added to the TS simulation formulation without updating the network admittance matrix, which is usually not accessible by users. The positive-sequence equivalent loads are equal to the three-phase total loads. One potential issue is that loads may not be adequate for representing the fault current contributions of the "active" ditribution systems to the transmission system.

*3) Multi-area Thevenin equivalent (MATE) for T&D system:* This interface model is different from the other interface models in terms of the equivalent models not being directly added to or used to update the system model in any simulator. The MATE is linked by the interface branches between the T&D system and the resulting link subsystem is solved outside the T&D simulators, which is illustrated in Fig. 2. Because the Thévenin equivalents are used outside the transmission simulator, use of Thévenin equivalents for representing the distribution systems become possible. The T&D dynamic co-simulation implementation based on this interface model combined with the parallel, iterative IS has been benchmarked against full EMT simulation in [2]. Thus, this IM will be used as the reference for comparing other interface models and interaction schemes.

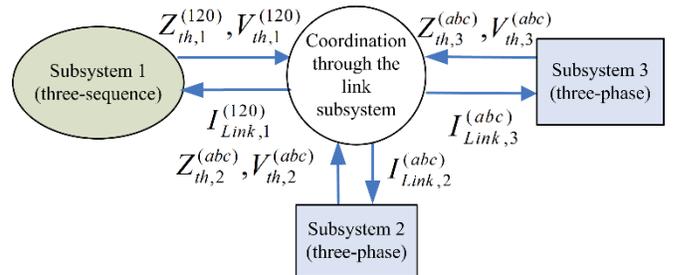

Fig. 2 Coordination of the subsystem simulation solutions through the link subsystem using the MATE approach [2]

### C. Interaction Schemes

ISs can be categorized into two categories including iterative and non-iterative, as shown in Table II. This categorization reflects the fact that some of the transmission and/or distribution system simulators do not provide the necessary access and functions to external users to allow inter-iteration operations for co-simulation at each time step. Both types of interaction schemes are compared. For each type, both series and parallel IMs are analyzed. For non-iterative type ISs, it is expected that the sequence of running the simulators may influence the co-simulation performance. In this paper, ISs 2 and 5 and the parallel scheme for the MATE as IM will be tested. Testing of other schemes will be reported in future publications.

TABLE II. Interaction Schemes for T&D Dynamic Co-Simulation

| # | Iteration or not | Interaction | |
|---|---|---|---|
| 1 | Non-iterative | Series | Transmission first |
| 2 | Non-iterative | Series | Distribution first |
| 3 | Non-iterative | Parallel | |
| 4 | Iterative | Series | Transmission first |
| 5 | Iterative | Series | Distribution first |
| 6 | Iterative | Parallel | |

## III. SIMULATION RESULTS

### A. Test System

A T&D test system (shown in Fig. 3) consisting of IEEE 9-bus system and three feeders that replace three loads in the transmission system is used for testing the interface techniques. Details of the test system can be found in [2]. For each load connected to the feeders, the load composition is 25% single-phase residential air-conditioner motor combined with 75% constant impedances. A three-phase-to-ground fault was applied to bus 5 in the transmission system and was cleared after 0.07 s. The simulation time step for T&D simulation is 0.005 s.

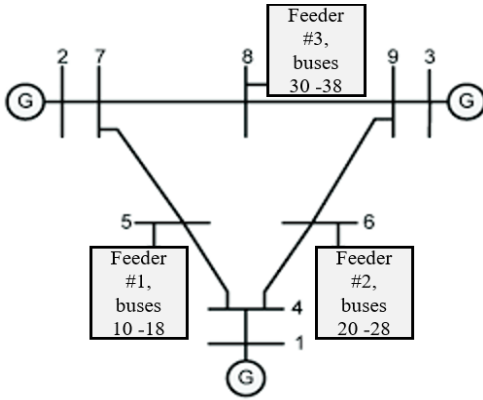

Fig. 3 A T&D test system

In the base case, the three-phase loads are balanced. To consider load unbalance conditions, a load unbalance factor $\beta$ is defined. Loads on the three phases are defined as:

$$P_{L,i}^A = \tfrac{1}{3} P_{Total,i},\ P_{L,i}^B = \tfrac{1-\beta}{3} P_{Total,i},\ P_{L,i}^C = \tfrac{1+\beta}{3} P_{Total,i} \quad (1)$$

where $P_{L,i}^A$, $P_{L,i}^B$ and $P_{L,i}^C$ are the real part of load on phases $A$, $B$ and $C$ of bus $i$, respectively, and $P_{Total,i}$ is the sum of loads.

The simulations were performed on a simulation platform developed based on InterPSS [7], which has been extended to support transmission (for both positive-sequence and three-sequence modeling) and distribution system dynamic simulation in [2]. Interface models shown in Table I have been developed on this platform for the comparison studies. Simulation results in terms of robustness, accuracy and efficiency will be discussed in the following subsections.

### B. Robustness

Network solution divergence was observed in the test cases for IMs 1 and 4, where forward-backward-sweep (FBS) power flow is used for the network solution of the distribution system DS. Divergence occurred at the fault clearing time step. As the voltage along the feeders sharply increased, the power consumption of stalled A/C motors increased proportionally to the square of terminal voltage. The high loading condition caused the power flow to diverge. In light of this, these two interface models will not be included in the following comparison studies. No network solution divergence issue was detected for the other IMs in this study.

### C. Accuracy of Different Interface Techniques

Using reference simulation results that are obtained with the MATE as IM and iterative parallel IS, the maximum and mean absolute errors of the positive sequence voltage magnitude at bus 5 ($\Delta V_5^+$), phase A voltage magnitude at the feeder bus 14 ($\Delta V_{14}^A$) and speed of the generator at bus 1 ($\Delta \omega_{G1}$), are used to measure the accuracy of different interfaced techniques. The comparison results are summarized in Table III. There are several important observations from the results: 1) For the same time step of 0.005 s, compared to non-iterative IMs, iterative IMs helped significantly improve the accuracy (from several times to tens of times) for the same interface model. The results justified the support of iteration in the development of HELICS [6]; 2) With the iterative IMs, representing the transmission system as *Thévenin equivalents* (IMs 3, 6, 7) in distribution system simulation in general produced higher accuracy results than the *voltage sources* (IMs 2, 5); 3) As the loading unbalance factor $\beta$ increased, the accuracy of IMs 2 and 3 dramatically decreased when iterative ISs were employed. In contrast, the accuracy of IMs 5 and 6 were much less impacted by the distribution system unbalanced conditions, mainly because the impact of unbalanced conditions on the transmission system can be better captured by the three-sequence networks. Thus, it is desirable to extend the existing positive-sequence modeling and TS simulation to three-sequence modeling and TS simulation for T&D dynamic co-simulation; and 4) For the same time step of 0.005 s, the accuracy of all tested IMs for non-iterative IS scenarios are not satisfactory. Improvement measures is needed.

### D. Computational Efficiency

The average computational time with different interface techniques for 15-second simulation is shown in Fig. 4. With the iterative scheme, the MATE based approach (IM # 7) showed much better efficiency than the others, which suggests that it has better convergence characteristic and can converge with fewer iterations. However, for non-iterative schemes, these IMs showed similar computational efficiency.

## IV. CONCLUSIONS AND FUTURE WORK

A comparative study of different interface techniques for T&D dynamic co-simulation was conducted in this paper. Firstly, the comparison results suggested that selecting the adequate transmission and distribution system simulators and iteration schemes should be a priority when developing T&D dynamic co-simulation. Secondly, three-sequence modeling for transmission systems is recommended to achieve better accuracy. Thirdly, the interaction schemes have significant impact on all the studied interface models. Iterative schemes should be used for better accuracy whenever possible. Lastly, for interface models, representing transmission systems as Thévenin equivalents is a more accurate choice in capturing transmission systems' impact on distribution systems.

Table III. Accuracy of Different Interface Models and Interaction Schemes

| Load unbalanced factor | Accuracy Metrics | Interface Techniques | | | | | | | | | |
|---|---|---|---|---|---|---|---|---|---|---|---|
| | | IS #5: Iterative, Series, Distribution first | | | | IS #6 | IS #2: Non-iterative, Series, Distribution first | | | | IS #3 |
| | | IM #2 | IM #3 | IM #5 | IM #6 | IM #7 | IM #2 | IM #3 | IM #5 | IM #6 | IM #7 |
| $\beta = 0$ | max $\Delta V_5^+$ | 2.1E-03 | 1.1E-03 | 3.1E-04 | 2.1E-04 | —— | 8.7E-02 | 2.9E-02 | 8.7E-02 | 6.2E-02 | 3.0E-02 |
| | avg. $\Delta V_5^+$ | 1.2E-03 | 3.5E-04 | 4.5E-05 | 4.6E-05 | —— | 2.8E-02 | 5.8E-03 | 3.4E-02 | 1.1E-02 | 4.0E-03 |
| | max $\Delta V_{14}^A$ | 3.5E-03 | 2.0E-03 | 4.7E-04 | 6.7E-04 | —— | 2.0E-01 | 2.4E-01 | 2.0E-01 | 2.4E-01 | 8.0E-02 |
| | avg. $\Delta V_{14}^A$ | 9.3E-04 | 1.0E-03 | 1.9E-04 | 1.6E-04 | —— | 3.7E-02 | 1.3E-02 | 3.4E-02 | 1.1E-02 | 8.0E-03 |
| | max $\Delta w_{G1}$ | 1.2E-04 | 2.0E-05 | 5.0E-05 | 5.0E-05 | —— | 1.8E-03 | 6.3E-04 | 1.8E-03 | 5.4E-04 | 1.0E-03 |
| | avg. $\Delta w_{G1}$ | 3.3E-05 | 4.0E-06 | 5.0E-06 | 9.0E-06 | —— | 6.8E-04 | 2.1E-04 | 6.8E-04 | 1.7E-04 | 1.9E-04 |
| $\beta = 10\%$ | max $\Delta V_5^+$ | 1.0E-02 | 8.0E-03 | 6.3E-04 | 3.7E-04 | —— | 8.3E-02 | 3.6E-02 | 8.3E-02 | 6.3E-02 | 5.9E-02 |
| | avg. $\Delta V_5^+$ | 2.9E-03 | 2.8E-03 | 4.0E-05 | 4.4E-05 | —— | 2.9E-02 | 2.0E-02 | 2.5E-02 | 2.1E-02 | 1.5E-02 |
| | max $\Delta V_{14}^A$ | 7.0E-02 | 3.6E-02 | 3.0E-03 | 4.5E-04 | —— | 1.9E-01 | 2.4E-01 | 2.1E-01 | 2.4E-01 | 2.6E-01 |
| | avg. $\Delta V_{14}^A$ | 2.4E-02 | 1.4E-02 | 1.6E-04 | 1.6E-04 | —— | 5.8E-02 | 3.7E-02 | 2.7E-02 | 2.5E-02 | 1.7E-02 |
| | max $\Delta w_{G1}$ | 3.6E-04 | 2.6E-04 | 4.0E-05 | 4.0E-05 | —— | 2.1E-03 | 1.5E-03 | 1.8E-03 | 1.5E-03 | 1.1E-03 |
| | avg. $\Delta w_{G1}$ | 1.9E-04 | 1.3E-04 | 7.0E-06 | 5.0E-06 | —— | 8.4E-04 | 6.0E-04 | 6.5E-04 | 5.3E-04 | 4.5E-04 |
| $\beta = 20\%$ | max $\Delta V_5^+$ | 2.7E-02 | 1.5E-02 | 1.6E-03 | 8.5E-04 | —— | 8.5E-02 | 3.8E-02 | 8.0E-02 | 6.5E-02 | 6.2E-02 |
| | avg. $\Delta V_5^+$ | 6.0E-03 | 5.3E-03 | 2.1E-04 | 2.2E-04 | —— | 3.4E-02 | 2.6E-02 | 2.3E-02 | 2.2E-02 | 1.8E-02 |
| | max $\Delta V_{14}^A$ | 1.3E-01 | 5.7E-02 | 7.3E-03 | 1.8E-03 | —— | 1.9E-01 | 2.3E-01 | 2.2E-01 | 2.3E-01 | 2.6E-01 |
| | avg. $\Delta V_{14}^A$ | 4.8E-02 | 2.1E-02 | 1.9E-03 | 2.8E-04 | —— | 8.5E-02 | 5.4E-02 | 3.1E-02 | 3.2E-02 | 2.2E-02 |
| | max $\Delta w_{G1}$ | 1.0E-03 | 5.8E-04 | 5.0E-05 | 4.0E-05 | —— | 2.7E-03 | 2.1E-03 | 1.8E-03 | 1.8E-03 | 1.5E-03 |
| | avg. $\Delta w_{G1}$ | 5.6E-04 | 2.2E-04 | 1.2E-05 | 1.0E-05 | —— | 1.3E-03 | 9.7E-04 | 7.2E-04 | 6.8E-04 | 6.8E-04 |

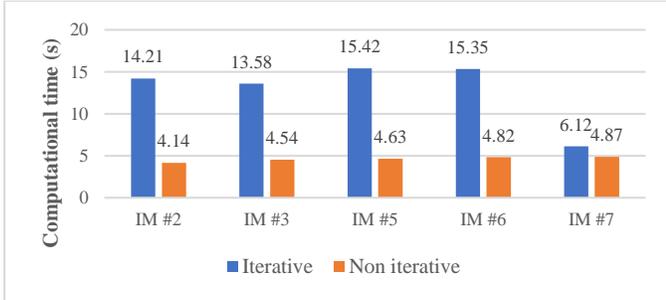

Fig. 4 Computational time with different interface techniques

Serval directions for future work are suggested: 1) the performance of different interface techniques under other fault conditions. 2) impact of different penetration levels of inverter-interfaced DER on performance of the interface models; 3) interface techniques suitable for distribution systems with non-radial topology (e.g., loop or network); 4) impact of the choice of different load models for representing distribution systems in the transmission system on the accuracy and robustness.